\documentclass[preprint]{aastex}

\slugcomment{May 4, 2000}

\shorttitle{Limit-Cycle Model of Star Formation}
\shortauthors{Hirashita \& Kamaya}

\begin{document}

\title{\bf APPLICATION OF THE LIMIT-CYCLE MODEL TO STAR FORMATION
HISTORIES IN SPIRAL GALAXIES: VARIATION AMONG MORPHOLOGICAL
TYPES}

\author{\bf HIROYUKI HIRASHITA\altaffilmark{1} AND
HIDEYUKI KAMAYA}
\affil{Department of Astronomy, Faculty of Science, Kyoto University,
Sakyo-ku, Kyoto 606-8502, Japan}

\email{hirasita@kusastro.kyoto-u.ac.jp}

\altaffiltext{1}{Research Fellow of the Japan Society for the Promotion of
Science.}

\begin{abstract}

We propose a limit-cycle scenario of star formation history for
any morphological type of spiral galaxies. It is known
observationally that the early-type spiral sample has a wider range
of the present star formation rate (SFR) than the late-type sample.
This tendency is understood in the framework of the limit-cycle
model of the interstellar medium (ISM), in which the SFR cyclically
changes in accordance with the temporal variation of the mass
fraction of the three ISM components. When the limit-cycle
model of the ISM is applied, the amplitude of variation of the SFR
is expected to change with the supernova (SN) rate. Observational
evidence indicates that the early-type spiral galaxies show
smaller rates of present SN than late-type ones. Combining this
evidence with the limit-cycle model of the ISM, we predict that the 
early-type spiral galaxies
show larger amplitudes in their SFR variation than the late-types.
Indeed, this prediction is consistent with the observed wider range
of the SFR in the early-type sample than in the late-type sample.
Thus, in the framework of the limit-cycle model of the ISM, we
are able to interpret
the difference in the amplitude of SFR
variation among the morphological classes of spiral galaxies.

\end{abstract}

\keywords{
galaxies: evolution ---
galaxies: ISM --- galaxies: spiral --- ISM: evolution ---
stars: formation} 

\section{INTRODUCTION}

In this paper, we propose a scenario of star formation histories
in giant spiral galaxies. Our standpoint is based on a very
interesting observational result
presented in Kennicutt, Tamblyn, \& Congdon (1994, hereafter KTC).
According to their sample galaxies, there is a difference in
present star-formation activity among morphological
types of spiral galaxies. This is found in KTC (the third paragraph 
of their \S 5.2). The early-type spiral galaxies have a
one-order-of-magnitude range in $b$, which denotes the 
present-to-past ratio of star-formation rate (SFR),
as being $b=0.01$--0.1. On the other hand, $b=0.5$--2, a range of just
a small factor, in the late-type spiral sample.
Moreover, in the framework of the scenario proposed in this paper,
we can incorporate another relevant observational result; the
difference of supernova  (SN) rate 
among the galactic morphology (Cappellaro et al.\ 1993).

Not only KTC, but also
Tomita, Tomita, \& Sait\={o} (1996)
found a difference in
present star-formation activity among different morphological
types of spiral galaxies (see also Devereux \& Hameed 1997).
Tomita et al.\ (1996) commented that this variation may be
a short-term change in the SFR in spiral galaxies and the
duration of an episode of star formation activity is
less than $10^8$ yr.

Kamaya \& Takeuchi (1997, hereafter KT97) independently put
an observational interpretation about the difference in
present star-formation activity among morphological types.
They pointed out that the short duration of star formation
proposed by Tomita et al.\ (1996) may indicate
that the interstellar medium (ISM) in a spiral galaxy
is a non-linear open-system. That is, they suggested that the
duration may result from the period of a limit cycle of mass
exchange among various phases of the ISM (see
Ikeuchi \& Tomita 1983, hereafter IT83, for the limit-cycle
behavior). If the ISM in a galaxy is regarded globally to be a 
non-linear open-system,
the evolution of whole ISM on a galaxy-wide scale may result
in the limit-cycle star formation history.
Indeed, the variance of the star formation activities of spiral
galaxies can be understood as a short-term ($\la 10^8$ yr)
variation of their activities. Recent result by Rocha-Pinto
et al.\ (2000) suggests that the Galactic star formation
history indeed shows such a galaxy-wide variability.

KT97's discussion was based on the sample of Tomita et al.\ (1996).
In the subsequent discussions, we re-examine KT97's
proposal more quantitatively
than their original considerations through a proper comparison with
KTC. Moreover, we check whether their scenario is consistent with
the difference of SN rate among the morphological types of spirals,
since the amplitude of the limit-cycle orbit is determined by
the SN rate (\S\ref{sec:scenario}). 
Thus, we can state clearly our motivation here.
We aim to interpret the difference in star formation history among 
morphological types in the framework of the limit-cycle model of
ISM in spiral galaxies.

According to a review by Ikeuchi (1988), he and his collaborators
indicated that we might understand the dynamical evolution of the ISM
on a galaxy-wide scale if we could describe galaxies as nonlinear
open systems (see also Nozakura \& Ikeuchi 1984, 1988).
Here, we stress that one type of their models, the limit-cycle model,
indicates a periodic star formation history
(KT97). A more elaborate model
is also proposed by Tainaka, Fukazawa, \& Mineshige (1993). 
To understand the behavior of star formation activity of spiral
galaxies, we focus on
this interesting hypothetical behavior of the ISM. Adopting the
limit-cycle model, KT97 insisted that
the amplitude of the cyclic part of the SFR, $\Psi'$, should be larger
than the SFR of the quiescent era, $\bar{\Psi}$. Moreover, adopting
a Schmidt (1959) law of index of 2, they predicted that the
amplitude ratio of $\Psi'$s, defined as
${\rm Max}(\Psi ')/{\rm Min}(\Psi ')$, should be $\sim 50$. Here, 
Max($\Psi '$) and
Min($\Psi '$) indicate the maximum and minimum values of $\Psi '$,
respectively. 
However, since ${\rm Max}(\Psi ')/{\rm Min}(\Psi ')$ depends on the
characteristic parameters for the limit-cycle model (IT83; or
\S\ref{sec:cyclic} in this
paper), we re-examine the amplitude for various parameters in this
paper.

Throughout this paper, thus, we discuss the difference in the
amplitude of cyclic star formation among the three morphological 
types of spiral galaxies (Sa, Sb, and Sc) along with the limit-cycle
model. In the next section, we review the
limit-cycle model of the ISM. In \S 3, we estimate the amplitude of 
cyclic SFR, and interpret the difference in the amplitude 
via the limit-cycle model. In \S 4, a consistent scenario for the 
time variation of SFR is proposed. In the final section,
we summarize our considerations and present some discussions.

\section{CYCLIC STAR FORMATION HISTORY}\label{sec:cyclic}

Since our discussions are based on mainly KTC in this paper,
we summarize firstly KTC.  KT97 is also
reviewed. Although KT97 discussed Tomita et al.\ (1996), their 
argument on the duration and the behavior of star formation
activity is not altered even if we are based on KTC.

\subsection{KTC's Sample and KT97's Interpretation}

Treating a data set of H$\alpha$ equivalent widths of galactic
disks with various morphologies, KTC has shown that the
star formation activities present a wide spread for each
morphological type (KTC's Fig.\ 6). They derived $b$ parameter
which indicates the ratio of the present SFR to the past-averaged
SFR.

Based on KT97, we re-interpret
Figure 6 in KTC: The KTC's wide dispersion of the star formation
activity is interpreted as evidence for a 
periodic star-formation history on the scale of a giant galaxy.
If galaxies have the same morphological type and the same age, such a
large scatter as that in Figure 6 of KTC should not appear for
near constant or monotonically declining SFRs. However, if
cyclic star formation occurs in any spiral galaxy,
we can easily understand why such a large scatter emerges. 
If the period of the cyclic star formation is several times
$10^7$ years, 
the dispersions in KTC's Figure 6 do not contradict the hypothesis
of KT97 by setting $b=\Psi '/\bar{\Psi}$, where $\Psi '$ and 
$\bar{\Psi}$ are the cyclic and past-averaged SFR, respectively.

Indeed, such a periodic star formation history is proposed
by Ikeuchi (1988) as cited by KT97.
His discussion is based on the limit-cycle behavior of the
fractional mass of each ISM shown by IT83.
If the fractional component of
the cold gas, where stars are formed, cyclically changes
on a short timescale ($\sim 10^7$--$10^8$ yr),
the SFR also varies cyclically.
Thus, we review the formulation by IT83 in the next subsection.

\subsection{Limit-Cycle Model of ISM}

We review the limit-cycle model for the ISM proposed by IT83 (see also
Scalo \& Struck-Marcell 1986). The model has been utilized to interpret
Tomita et al.\ (1996) (KT97). The
limit-cycle behavior
emerges if we treat the ISM as a non-linear open system.
As long as the ISM is a non-linear open system, 
it spontaneously presents a dissipative structure
(Nozakura \& Ikeuchi 1984).

First of all, we should note that the galaxy disk is treated as one zone.
The interstellar medium is assumed to consist of three components
each with its temperature $T$ and density $n$
(McKee \& Ostriker 1977); the hot rarefied gas
($T\sim 10^6$ K, $n\sim 10^{-3}$ cm$^{-3}$), the warm gas
($T\sim 10^4$ K, $n\sim 10^{-1}$ cm$^{-3}$), and the cold
clouds ($T\sim 10^2$ K, $n\sim 10$ cm$^{-3}$).
The fractional masses of the three components are denoted by
$X_{\rm h}$, $X_{\rm w}$, and $X_{\rm c}$, respectively.
A trivial relation is
\begin{eqnarray}
X_{\rm h}+X_{\rm w}+X_{\rm c}=1.\label{trivial}
\end{eqnarray}

The following three processes are considered in IT83
(see also Habe, Ikeuchi, \& Tanaka 1981):
[1] the sweeping of the warm gas into the cold component at the
rate of $a_*X_{\rm w}$ ($a_*\sim 5\times 10^{-8}$ yr$^{-1}$);
[2] the evaporation of cold clouds
embedded in the hot gas at the rate of $b_*X_{\rm c}X_{\rm h}^2$
($b_*\sim 10^{-7}$--$10^{-8}$ yr$^{-1}$);
[3] the radiative cooling of the hot gas by mixing with the ambient
warm gas at the rate of $c_*X_{\rm w}X_{\rm h}$
($c_*\sim 10^{-6}$--$10^{-7}$ yr$^{-1}$). Writing down the rate
equations and using equation (\ref{trivial}), IT83 obtained
\begin{eqnarray}
\frac{dX_{\rm c}}{d\tau} & = & -BX_{\rm c}X_{\rm h}^2+
A(1-X_{\rm c}-X_{\rm h}),\label{eq:cold} \\
\frac{dX_{\rm h}}{d\tau} & = & -X_{\rm h}
(1-X_{\rm c}-X_{\rm h})+BX_{\rm c}X_{\rm h}^2,
\label{eq:hot}
\end{eqnarray}
where $\tau\equiv c_*t$, $A\equiv a_*/c_*$, and $B\equiv b_*/c_*$.

The solutions of equations (\ref{eq:cold}) and (\ref{eq:hot}) are
classified into the following three types (IT83):

[1] $A>1$; all the orbits in the $(X_{\rm c},\; X_{\rm h})$-plane
reduce to the node (0, 1) (node type),

[2] $A<1$ and $B>B_{\rm cr}$; all the orbits reduce to a stable
focus $[(1-A)/(AB+1),\; A]$ (focus type),

[3] $A<1$ and $B<B_{\rm cr}$; all the orbits converge on a
limit-cycle orbit (limit-cycle type),

\noindent
where $B_{\rm cr}\equiv (1-2A)/A^2$.
Obviously, case [3] is important if we wish to predict a cyclic star
formation history.
According to the summary of the limit-cycle model by Ikeuchi (1988), 
the period of a cycle is several times $10^7$ years, as depicted in 
his Figure 4. Since
this period is much smaller than the characteristic timescales in 
galaxy evolution such as the gas consumption timescale
($>1$ Gyr: KTC's $\tau_{\rm R}$), the cyclic change of SFR will
produce a scatter in the observed star formation activities in 
spiral galaxies even if their ages are similar.

\section{OSCILLAROTY MODEL OF SFR}

\subsection{Model Description}

According to KT97, we use a simple description
to test our discussion.
First, we define the present quiescent component of the SFR as
$\bar{\Psi}$ and the oscillatory component of the SFR as $\Psi '$. 
Then, the total SFR is denoted as
\begin{eqnarray}  
\Psi = \bar{\Psi} + \Psi ' .
\end{eqnarray}
Our definitions are adequate when the period of oscillation of the SFR
is much smaller than the cosmic age (e.g., Sandage 1986).
According to Schmidt (1959), the SFR in a galaxy is approximately
expressed as ${\rm SFR} \propto n^p$ ($1<p<2$), where $n$ is the
mean gas density of the galaxy. If we interpret $n$ as the
gas density of a cold cloud, which can contribute to the
star formation activity,  we expect the oscillatory
part of the SFR to be
\begin{eqnarray}
\Psi ' \propto X_{\rm c}^{1.5},\label{schmidt}
\end{eqnarray}
where we have assumed that $p=1.5$ (Kennicutt 1998). Using this 
relation and equations (\ref{eq:cold}) and (\ref{eq:hot}), the
variation of the star formation activity (i.e., $\Psi '$
as a function of $\tau$) is calculated.
For example, according to Figure 6 in Ikeuchi (1988), 
this mass fraction of the cold gas, $X_{\rm c}$, can vary from
$\sim 0.1$ to
$\sim 0.9$. Thus, we expect the magnitude of the variation of 
$\Psi '$ to be about two orders of magnitude during one period of
the oscillation.

Since our model follows IT83, the structure of a model galaxy is
hypothesized to be one-zone, 
that is, the local phenomena of the ISM are averaged in space. 
The simplicity of the one-zone approximation gives the advantage 
that the background
physical processes are easy to see. Moreover, even though the effect
treated in this paper and IT83 may be local, a ``global'' (i.e.,
galactic-scale) dissipative structure of the ISM emerges
(Nozakura \& Ikeuchi 1984), as long as a galaxy is assumed to be
a non-linear open system (Nicolis \& Prigogine 1977). Thus, as a first
step, we treat a model galaxy as
being a one-zone object which is a non-linear open system.

Habe et al.\ (1981) stated in their \S 7 that for the one-zone
assumption to be acceptable it is necessary that the mean distance 
between supernova remnants (SNRs) be less than 100 pc
(if a characteristic lifetime of SNRs of 
$\tau_{\rm life}\sim 10^7$ yr and a mean
expansion velocity of 10 km s$^{-1}$ are adopted). This is because
the SNRs should affect the whole disk for the one-zone treatment.
The distance of less than 100 pc means that there are $N\sim 10^4$
SNRs in a galaxy disk, if the disk size of 10 kpc is assumed.
This number is possible if SNe occur every $10^3$ yr
($\tau_{\rm life}/N\sim 10^7$ [yr]/$10^4$). Considering that the SN
rate in a spiral galaxy is typically 1/100--1/50 yr$^{-1}$
(Cappellaro et al.\ 1993), the mean distance between
SNRs is less than 100 pc even if 10--20 massive stars are clustered
in a region.

Indeed, from the observational viewpoint, Rocha-Pinto et al.\
(2000) have shown that the star formation history of the
Galaxy does present a short-term variability whose timescale
is less than $\sim 1$ Gyr (see also Takeuchi \& Hirashita 2000).
Though they adopted the sample stars in the solar neighborhood,
they showed by
estimating the diffusion time of the stars that the sample represents 
the stars in the Galaxy-wide scale. Based on this 
observational evidence as well as the discussions in the
previous two paragraphs, we accept the one-zone treatment by 
IT83 and apply it to the ISM on a galactic-wide
scale.

Once we accept the one-zone treatment, we need global observational
measures of a galactic disk to examine our scenario. In this paper,
the H$\alpha$ equivalent width in KTC is the global physical
parameter. KTC declared that their data excluded the bulge component 
and that
the disk component is uncontaminated. 

\subsection{Application to KTC}

For the comparison between the model prediction and the observational
data, we relate $b$ defined in KTC
(the ratio of the present SFR to the past-averaged SFR) to
the model prediction.  The parameter $b$
is calculated from the equivalent width of H$\alpha$ emission.
According to KT97, we can assume that
$b \simeq\Psi '/\bar{\Psi}$ if the large variance of
$b$ in Figure 6 of KTC originates from a short-term
variation. We combine the IT83's model with the star
formation history via the Schmidt law
(equation \ref{schmidt}). For example, when
$X_{\rm c}=0.1$ at the minimum SFR and $X_{\rm c}=0.7$
at the maximum (Fig.\ 1 of IT83), the value of $X_{\rm c}^p$ changes
from 0.03 to 0.59 during the cycle if $p=1.5$. 
Accordingly, we find that the maximum SFR is about 20 times larger 
than the minimum SFR, since $\Psi '$ is proportional to
$X_{\rm c}^p$ (equation \ref{schmidt}).
Thus, using the cyclic star-formation scenario, we find
the maximum of  
$b$ also becomes 20 times larger than the minimum $b$
in this numerical example.

To summarize, the large variance of $b$ in Figure 6 of
KTC is naturally derived through the Schmidt law,
if the limit cycle model is a real evolutionary picture of ISM. In
the next section, we examine this
point more precisely, in order to reproduce the variance of
star formation activities for each morphological type of spiral
galaxies. In the following discussions, we examine
Max($\Psi'$)/Min($\Psi'$), where Max($\Psi'$) and Min($\Psi'$) are
maximum and minimum values of the oscillatory SFR
(the maximum and minimum are defined by
the maximum and minimum star formation rates during a
period of the limit cycle, respectively), and thus
${\rm Max}(\Psi' )/{\rm Min}(\Psi' )\equiv
{\rm Max}(X_{\rm c}^{p})/{\rm Min}(X_{\rm c}^{p})$. 
In the rough estimate in the previous paragraph,
${\rm Max}(\Psi' )/{\rm Min}(\Psi' )$ is 20 with $p=1.5$. Here, we 
define 
\begin{eqnarray}
F_{\rm c} \equiv \frac{ {\rm Max}(\Psi ') }
                      { {\rm Min}(\Psi ')},\label{def_fo} 
\end{eqnarray}
for convenience in the subsequent sections.
Using this relation and equations (\ref{eq:cold}) and
(\ref{eq:hot}), the variation of the star formation activity
(i.e., $\Psi '$ as a function of $\tau$) is calculated, and
$F_{\rm c}$ is evaluated finally.

\section{SCENARIO OF LIMIT-CYCLE STAR FORMATION}\label{sec:scenario}

To propose a scenario of star formation 
history for spiral galaxies based on the limit-cycle model, let us
start with a very interesting observational result.
According to KTC, the early-type sample has a larger variance
of SFR than the late-type sample. 
As a first step, we reconstruct this observational tendency 
in the framework of IT83. 
Then, we perform several numerical analyses to examine parameters
which implement the limit-cycle oscillation of the cold phase of the 
ISM. For the ISM in spiral galaxies,
the  full possible ranges of the parameters of $A$ and $B$ 
(e.g., Habe et al.\ 1981) 
are $A=a_*/c_*$ of $\sim 0.05$ to $\sim 0.5$
and that $B=b_*/c_*$ of $\sim 0.01$ to $\sim 1$, respectively. 
In the following discussions, we focus on the parameter sets for the
limit-cycle type (case [3] in \S 2.2).

\begin{figure}[htb]
\figurenum{1}
\centering\includegraphics[width=10cm]{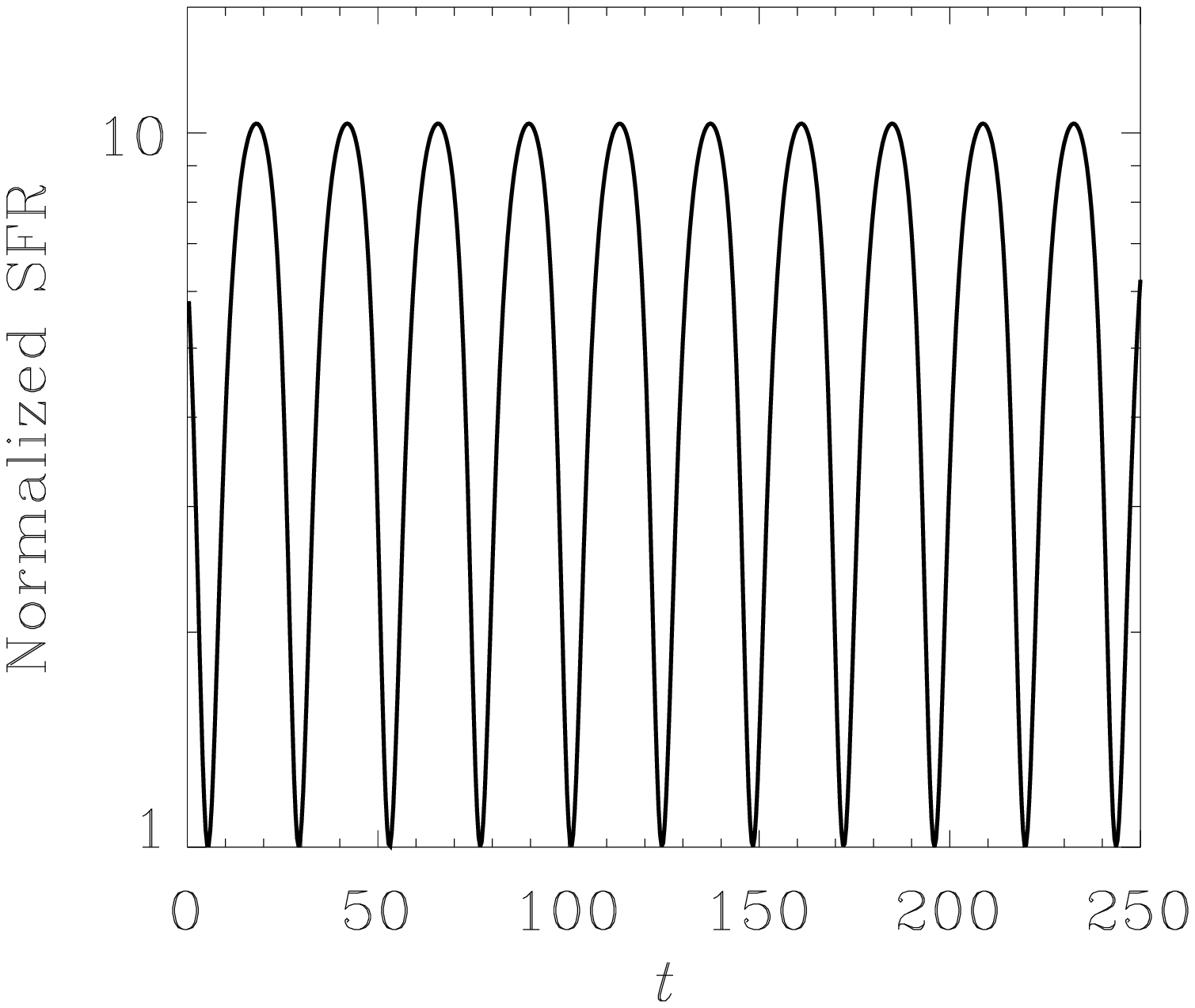}
\centering\includegraphics[width=10cm]{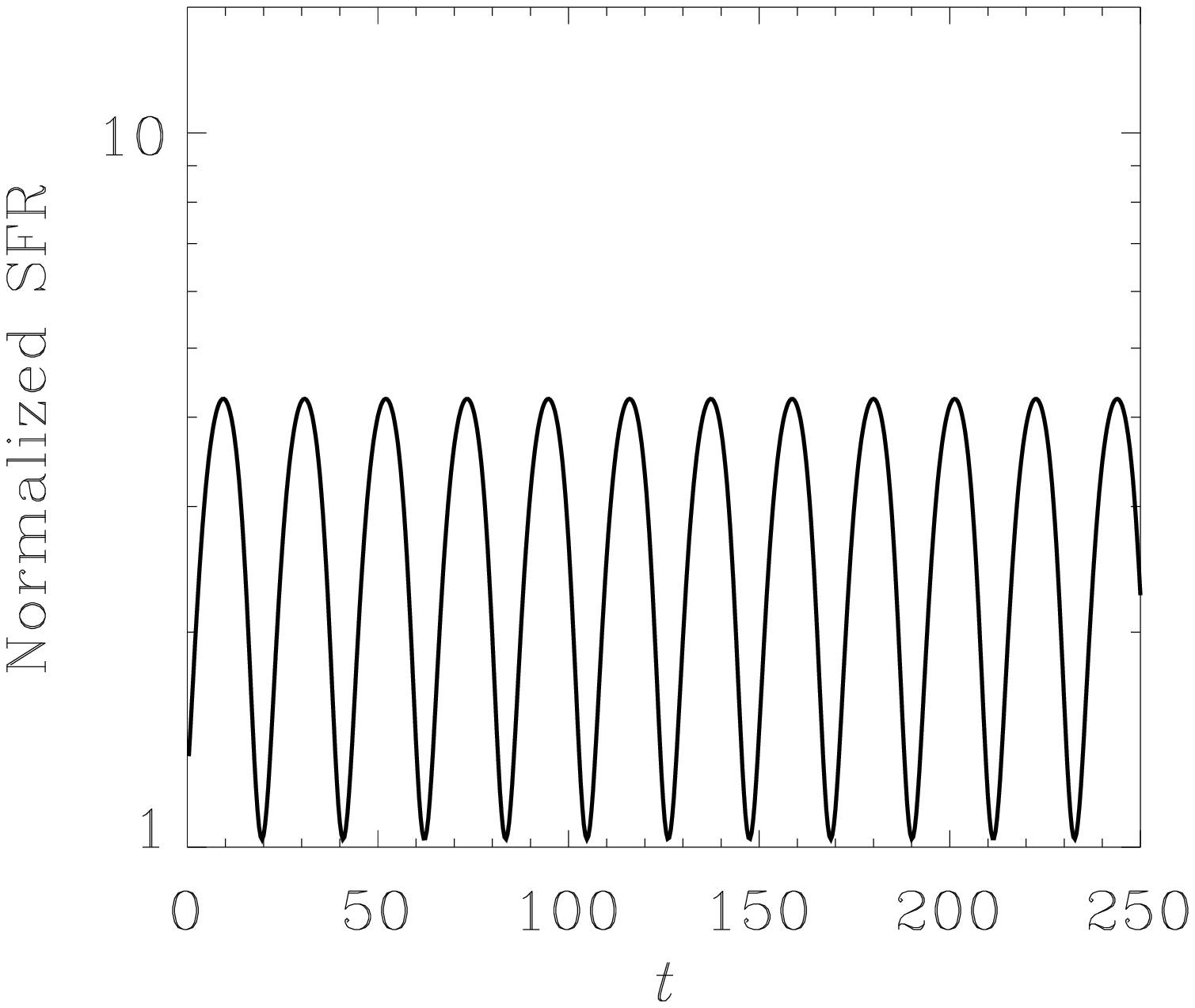}
\figcaption[fig1.ps]{Time variation of SFR normalized by the
minimum SFR. The timestep is normalized to the cooling time of the 
hot gas
determined by the rate of mixing between hot and warm
components ($\sim 10^6$ yr). ({\it a}) $A=0.38$ and $B=1.0$
are adopted to match the variance of the Sa sample in KTC. The 
amplitude $F_{\rm c}$ is
10. ({\it b}) $A=0.40$ and $B=1.0$ match the variance of the Sc
sample in KTC. The amplitude $F_{\rm c}$ is 4.
}\label{fig1} 
\end{figure}

Two of the results of differently parametrized limit-cycle
behavior are displayed in
Figures \ref{fig1}{\it a}--\ref{fig1}{\it b}, where the SFRs are
normalized to the minimum SFR. In Figure \ref{fig1}{\it a}, we
find an amplitude ($F_{\rm c}$) of about 10,
which might correspond to the result of the Sa sample in KTC.
Figure \ref{fig1}{\it b} corresponds to the amplitude of about 4 for
Sc in KTC.

Clearly, the difference in the amplitude between 
the Sa and Sc galaxies for the KTC data sets 
can be reproduced via the models which yield
Figures \ref{fig1}{\it a} and \ref{fig1}{\it b}. We also present
$F_{\rm c}$ for various $A$ and $B$ in Table 1, from which we
observe that the value of $F_{\rm c}$ is more sensitive to $A$ than
to $B$. Indeed, from the rough estimate, 
$\delta F_{\rm c}/\delta A \sim -133/0.08 \sim -1600$  for $B=1.0$ and
$\delta F_{\rm c}/\delta B \sim -119/1.5 \sim -80$ for $A=0.34$.
Thus, the two figures are presented for different values
of $A$. Here, we state an important point:
The early-type spiral galaxies favor a small $A$, 
while the late-type ones are consistent with a large $A$.

\begin{table}[t]
\small
\begin{center}
Table~1.\hspace{4pt}$F_{\rm c}$ as a function of $A$ and $B$.\\
\vspace{6pt}
\begin{tabular}{c|c|ccccc}
\hline\hline
\multicolumn{2}{c|}{} & \multicolumn{4}{c}{$B$} \\ \cline{3-6}
\multicolumn{2}{c|}{} & 0.5 & 1.0 & 1.5 & 2.0 \\
\hline
    & 0.32 & 355 & 137 & 61  & 29 \\
    & 0.34 & 129 & 52  & 24  & 10 \\
$A$ & 0.36 & 57  & 23  & 9   & 3 \\
    & 0.38 & 27  & 10  & 4   & --- \\
    & 0.40 & 15  & 4  & ---   & --- \\
[4pt] \hline
\end{tabular}
\end{center}
\tablecomments{We show the value of $F_{\rm c}$ if $B < B_{\rm c}$
(the condition for the limit-cycle behavior) is satisfied.
Otherwise, `---' is marked.
See text for the definitions of quantities.}

\end{table}

The variation of the amplitude in accordance with $A$ and $B$ is
qualitatively interpreted as follows. Small $A$ (or small $B$)
indicates that the transition from the warm to the hot component
(or the cold to the warm component) is inefficient. Thus,
when $A$ (or $B$) is small, we must wait for 
$X_{\rm w}$ (or $X_{\rm c}$) to become large before the phase
transition can become important, since the transition rate is
described by $AX_{\rm w}$ (or $BX_{\rm c}X_{\rm h}^2$). Thus,
the amplitude and the period become large
for small $A$ (or $B$). This interpretation of the relation
between $A$ (or $B$) and the amplitude is qualitatively robust. 
This means that  the scenario proposed in this paper is unchanged
even if a more elaborate model such as 
Ikeuchi, Habe, \& Tanaka (1984) is used.

As a next step, we examine the SFR variance via the effect of $A$.
According to the definition of $A$ in \S 2.2, we expect a larger
rate of SNe for large $A$ (e.g., Sc)
than for small $A$ (e.g., Sa).  Then, the result in the previous
paragraph predicts an important point that
the early-type spiral galaxies have smaller {\it present} SN rate
than the late-type spirals. 
This is confirmed in the following two points:

\noindent
[1] The higher SFR per unit optical luminosity in later
type spiral galaxies (Figure 6 of KTC) indicates that the Type II
SN rate per
unit optical luminosity is higher in later types. Thus, it 
is natural in the context of our model that the late-type spirals 
have larger $A$ than the early-types.

\noindent
[2] The expected trend of the SN rate for early-to-late types has
been found by Cappellaro et al.\ (1993).
They examine the SN rate per blue luminosity in various types of 
spirals, and present a summary of their results in their Table 4.
We can confirm via their Table 4 that 
our scenario of limit-cycle SFR is consistent with the
observational trend of SN rate as the galactic morphology varies. 
Moreover, the present SFR is reflected by the
present rate of
Type II SNe. According to Cappellaro et al., 
Sc-types show higher Type II SN rates than Sa types.
Then, we can infer that the SFR of Sc galaxies is larger than that of Sa
galaxies, which is compatible with Figure 6 of KTC.

\noindent
{}From these pieces of evidence, we find a consistent picture of
the SFR variance in the spiral sample as a function of
morphology via the scenario of the limit-cycle star formation
history. We note that the large gas-to-stars mass ratio in
late-type spiral galaxies is probably the reason for the large SFR,
and that a larger mean SFR yields a smaller variance because
of a larger SN rate.

The trend of $F_{\rm c}$ with varying $B$ is also consistent with
the different SN rates among morphological types. Since $B$
physically means the efficiency of the evaporation of the cold
component via conduction (one of the so-called SN feedback effects), 
$B$ increases with increasing
SN rate. Because a large value of $B$ tends to reduce $F_{\rm c}$
as can be seen in Table 1, a small $F_{\rm c}$ is caused
when the SN rate is large. Thus, from a similar argument to
that in the previous paragraph, late-type spiral galaxies ought to
have small values of $F_{\rm c}$.
Considering the sensitivity of $F_{\rm c}$ to $A$ (\S 4), however, 
we insist that $A$, not $B$, is the dominant contributor to the
determination of the amplitude $F_{\rm c}$.

To be fair, the picture presented in this paper may not be the 
unique interpretation of the variation of SFR. A stochastic
fluctuation may easily reproduce the observed scatter of
SFR in the KTC's sample as commented in \S 5.2.

\section{SUMMARY AND IMPLICATIONS}

\subsection{Summary}

In this paper, we have demonstrated that the large variety of SFR
in spiral sample in KTC may result from the limit-cycle evolution
of ISM as suggested by KT97. We present this more
quantitatively by using the numerical modeling by IT83 and explain 
the difference
in the range of SFR between morphological types.
It is known observationally that the early-type spiral sample
has a wider range of the present SFR than the late-type sample
(KTC). In our framework of the limit-cycle scenario of star-formation 
history in spiral galaxies, the early-types should show a more evident 
time variation of the SFR than the late-types to be  
consistent with the fact that Sc galaxies has the higher
present SN rate than Sa galaxies (\S\ref{sec:scenario}). Thus, 
the limit-cycle model by Ikeuchi and his collaborators provides
a consistent picture of the ISM evolution for any type of
spiral galaxies.  

\subsection{Implications}

What is the underlying physical mechanism responsible for the
variation in $A$ and $B$? 
Since $A$ and $B$ are related to the Type II SN rate, 
this question reaches the most basic and unsolved question:
What is the physical mechanism responsible for the different
star formation activities among morphological types?
First of all, recall that the later spirals have
larger bulge-to-disk ratio than the earlier spirals.
This can mean that the net volume of disk of the later spirals
is larger than that of the earlier spirals. Once we accept
the larger volume of disk of the later spiral galaxies,
we expect that Type II SN rate per galaxy is higher in the late types
than in the early types because of the large disk, where on-going
star formation is generally observed. Moreover, the parameter $c_*$
is determined from the mixing rate between the warm and hot gases.
Then, if the volume of disk is effectively larger in the late
types than in the early types, the late-type spirals may have
smaller values of $c_*$ than the early-types because the ISM
will travel a larger distance before the mixing. 
Since $A=a_*/c_*$ and $B=b_*/c_*$,
the late type spirals tend to have larger $A$ and $B$ than
the early types.
Therefore, as an implication, we propose that 
the size of the disk of galaxies  is a factor that physically
produces the difference in $A$ and $B$ among the morphologies of
spirals.

We expect that the difference in $A$ and $B$ is produced by
an interplay between the size effect described in the previous
paragraph and the SN rate as mentioned in \S 4.
In fact, the question has been answered from an observational
viewpoint in \S 5.3 of KTC by stating
``From an observational point of view, the progression in disk star
formation histories with morphological type is not surprising,
since one of the fundamental classification criteria is disk
resolution, which should relate at least indirectly to the
fraction of young stars in the disk.''

In this paper, we present a consistent picture for
the variance of star formation activities in spiral galaxies by
relating the differences in variance among morphological classes
with the supernova rate. However, since an earlier-type
sample has a
lower gas-to-stellar mass ratio, its mean SFR will be lower but its
variance will in any case tends to be larger because of the stochastic
fluctuations. 
In order to see whether the variance is caused by a purely stochastic
process or not, an amplitude of a stochastic SFR should be
given in a physically reasonable way. In other words, we should specify 
what kind
of the stochastic process is physically reasonable.
Though our model is not
stochastic, it provides a way to give an amplitude of the
variable SFR. To be fair, however, another modeling for the
variation of the SFR, probably a stochastic modeling, may
provide another interpretation for the variance of SFR in KTC.

Observational study is now progressing. Recently, 
Rocha-Pinto et al.\ (2000)
found that the star formation history of the Galaxy is indeed
oscillatory. Based on their work, Takeuchi \& Hirashita (2000)
proposed that the frequency distribution function of the SFR of
the Galaxy sampled every 0.4 Gyr from the formation of the
Galactic disk shows a flat distribution. Comparing this
distribution with the distribution of $b$ in KTC will offer us
a useful element in helping
to judge whether the scatter of $b$ is caused by an oscillatory
behavior of the SFR in any spiral galaxy.

\acknowledgements

We wish to thank the anonymous referee for invaluable comments
that substantially improved the discussion of the paper.
We are grateful to S. Mineshige for continuous encouragement.
We also thank A. Tomita and T. T. Takeuchi for useful discussions. 
One of us (HH) acknowledges the Research Fellowship of the Japan
Society for the Promotion of Science for Young Scientists. We
fully utilized the
NASA's Astrophysics Data System Abstract Service (ADS).


\begin{references}


Cappellaro, E., Turatto, M., Benetti, S., Tsvetkov, D. Yu.,
Bartunov, O. S., \& Makarova, I. N.\ 1993, A\&A, 273, 383


Devereux, N. A., \& Hameed, S. 1997, AJ, 113, 599


Habe, A., Ikeuchi, S., \& Tanaka, Y. D. 1981, {PASJ}, {33}, 23



Ikeuchi, S., 1988, Fundam.\ Cosmic Phys., {12}, 255

Ikeuchi, S., Habe, A., \& Tanaka, Y. D. 1984, MNRAS, 207, 909

Ikeuchi, S., \& Tomita, H. 1983, {PASJ}, {35}, 77 (IT83)

Kamaya, H., \& Takeuchi, T. T. 1997, {PASJ}, {49}, 471 (KT97)

Kennicutt, R. C. Jr. 1998, ApJ, 498, 541

Kennicutt, R. C. Jr., Tamblyn, P, \& Congdon, C. W. 1994, ApJ, 435,
22 (KTC)



McKee, C. F., \& Ostriker, J. P. 1977, ApJ, 218, 148

Nicolis, G., \& Prigogine, I. 1977, Self-Organization in Nonequilibrium
Systems (New York: Wiley and Sons)

Nozakura, T., \& Ikeuchi, S. 1984, {ApJ}, {279}, 40

Nozakura, T., \& Ikeuchi, S. 1988, {ApJ}, {333}, 68

Rocha-Pinto, H. J., Scalo, J., Machiel, W., \& Flynn, C. 2000, ApJ,
531, L115

Sandage, A. 1986, {A\&A}, {161}, 89

Scalo, J. M., \& Struck-Marcell, C. 1986, ApJ, 301, 77

Schmidt, M. 1959, {ApJ}, {129}, 243

Tainaka, K., Fukazawa, S., \& Mineshige, S. 1993, {PASJ}, {45}, 57

Takeuchi, T. T., \& Hirashita, H. 2000, ApJ, submitted

Tomita, A., Tomita, Y., \& Sait\={o}, M. 1996, {PASJ}, {48}, 285


\end{references}
\end{document}